\documentclass[journal]{IEEEtran}
\ifCLASSINFOpdf
\else
\fi
\usepackage[utf8]{inputenc}
\usepackage{graphicx}
\usepackage{hyperref}
\usepackage[section]{placeins}
\usepackage[section]{placeins}
\usepackage{lscape}

\begin{document}
%
\title{\LARGE{Unsupervised segmentation of biomedical hyperspectral image data: tackling high dimensionality with convolutional autoencoders}}
%
%
%

\author{Ciaran~Bench, Jayakrupakar~Nallala, Chun-Chin~Wang, Hannah~Sheridan, Nicholas~Stone \\ School of Physics and Astronomy, University of Exeter, Exeter, UK
}

\maketitle

\begin{abstract}
Information about the structure and composition of biopsy specimens can assist in disease monitoring and diagnosis. In principle, this can be acquired from Raman and infrared (IR) hyperspectral images (HSIs) that encode information about how a sample's constituent molecules are arranged in space. Each tissue section/component is defined by a unique combination of spatial and spectral features, but given the high dimensionality of HSI datasets, extracting and utilising them to segment images is non-trivial. Here, we show how networks based on deep convolutional autoencoders (CAEs) can perform this task in an end-to-end fashion by first detecting and compressing relevant features from patches of the HSI into low-dimensional latent vectors, and then performing a clustering step that groups patches containing similar spatio-spectral features together. We showcase the advantages of using this end-to-end spatio-spectral segmentation approach compared to i) the same spatio-spectral technique \emph{not} trained in an end-to-end manner, and ii) a method that only utilises spectral features (spectral k-means) using simulated HSIs of porcine tissue as test examples. Secondly, we describe the potential advantages/limitations of using three different CAE architectures: a generic 2D CAE, a generic 3D CAE, and a 2D CNN architecture inspired by the recently proposed UwU-net that is specialised for extracting features from HSI data. We assess their performance on IR HSIs of real colon samples. We find that all architectures are capable of producing segmentations that show good correspondence with HE stained adjacent tissue slices used as approximate ground truths, indicating the robustness of the CAE-driven spatio-spectral clustering approach for segmenting biomedical HSI data. Additionally, we stress the need for more accurate ground truth information to enable a precise comparison of the advantages offered by each architecture.
\end{abstract}


%
\IEEEpeerreviewmaketitle

\section{Introduction}
Images depicting spatially resolved structural and compositional information about a sample tissue can assist pathologists in disease monitoring and diagnosis \cite{alexandrov2013segmentation, lu2014medical,ul2021review,diem2013molecular}. A given tissue component or region is generally differentiated from others by i) its constituent molecules and ii) the way in which these are typically distributed in space. This information is encoded in a Raman or infrared (IR) hyperspectral image (HSI) (an image that can be formed by collating Raman or infrared spectra acquired at equally spaced out points across a sample), that reveals the presence/quantity of molecular bonds unique to specific molecules at each location \cite{scotte2018assessment,diem2013molecular}.


Spectral k-means is a common strategy for segmenting tissue components from Raman/IR HSIs \cite{khouj2018hyperspectral,hedegaard2010discriminating,hedegaard2011spectral,krafft2011crisp,piqueras2015combining}. In essence, pixels are grouped together based on the similarity of their constituent spectra. Provided a suitable number of cluster groups are chosen, the resultant cluster map (an image formed by displaying the cluster group assigned to each spectrum in the HSI) reveals where particular molecules reside in the sample, and can be used to segment a tissue into regions defined by their distinct molecular contents. But as stated previously, a given tissue component or region may also be defined by its unique \emph{arrangement} of molecules (i.e. how their constituent spectra are arranged in space). Spectral k-means is inadequate for segmenting tissue regions in this way as it does not use context of the spatial arrangements of neighbouring spectra. Instead, a more optimal approach would involve i) devising a method for detecting how spectra are arranged in various regions of the HSI and ii) finding a way to cluster regions containing similar arrangements of spectra. Though, the most effective way to detect and manipulate the spatial and spectral features unique to each sample component remains an open question. This is particularly challenging given the high dimensionality of HSI datasets that demand large amounts of memory and long computation times to process.

Recently, deep learning approaches have emerged as promising tools for segmentating HSI data \cite{li2019deep}. Compared to `classical' spatio-spectral segmentation approaches such as phasor analysis, extended morphological profiles, and sparse representation models \cite{fu2014reliable, tipping2022stimulated, ul2021review,manifold2021versatile,li2019deep, chen2011hyperspectral,fang2014spectral, camps2006composite,gorretta2012iterative,lu2014medical,baassou2012unsupervised}, or other generic hand-crafted feature extraction techniques \cite{li2018discriminatively} such as scale-invariant feature transform descriptors \cite{van2009evaluating}, local binary pattern operators \cite{guo2010completed}, non-negative matrix factorization \cite{hong2015joint}, and the complex wavelet structural similarity index \cite{sampat2009complex}, networks learn optimal feature extraction automatically and are therefore more readily applicable to problems where it may be challenging to hand-engineer desirable features and/or consistently extract them from the data. Several architectures have been used for this task, such as recurrent neural networks \cite{hang2019cascaded}, transformers \cite{hong2021spectralformer} (both extract long and short range spectral features by treating HSI datacubes as sequences), graph convolutional networks (effective at capturing long range spatial dependencies) \cite{hong2020graph}, and generative adversarial networks \cite{zhu2018generative} (improving network generalisability with an adversarial training scheme). However, architectures based on convolutional networks appear to be the most commonly implemented. All of these have produced promising results on various HSI datsets in supervised or semi-supervised schemes \cite{li2019deep,lee2017going,paoletti2018capsule,li2018classification, he2017generative,wu2017semi,li2014classification,midhun2014deep,chen2015spectral,wu2017semi,liu2017semi,kang2019semi,shahraki2020deep,fang2019hyperspectral,wambugu2021hyperspectral,mei2019unsupervised,manifold2021versatile,chen2016deep,chen2014deep}. However, in response to the usual lack of paired training data, effort has been placed towards developing fully unsupervised strategies.

Most reported methods for performing segmentation without ground truths utilise conventional analysis approaches as opposed to deep networks \cite{barbato2022unsupervised,schclar2017diffusion,ajay2022unsupervised,zhao2019fast,wang2015unsupervised,mughees2016unsupervised,ye2010segmentation,li2010hyperspectral,gillis2014hierarchical, murphy2018unsupervised}. However, one approach based on the use of fully convolutional autoencoders stands out in this respect \cite{nalepa2020unsupervised,obeid2021unsupervised,ajay2022unsupervised}. Here, the HSI is decomposed into patches and a convolutional autoencoder is used to detect and compress information about spatial and spectral features found in each patch into low-dimensional latent vectors to enable their clustering in an end-to-end fashion (therefore, each stage of the data processing is specifically optimised for the target segmentation task) \cite{nalepa2020unsupervised}. This technique belongs to a broader class of approaches that use supervisory signals from a secondary clustering step to enhance the quality of the learned latent representation via a unified learning objective \cite{li2018discriminatively,xie2016unsupervised,yang2016joint,liu2018infinite}. Fully convolutional autoencoders (as opposed to other architectures such as stacked autoencoders \cite{hinton2006reducing,vincent2010stacked,larochelle2009exploring,masci2011stacked,lee2011unsupervised, xie2016unsupervised}, or more generic autoencoders \cite{song2013auto,huang2014deep}) are well-suited to processing HSI data as they can easily manipulate structured/high dimensional image data, are specialised to extract spatial features, and have a straightforward training procedure. Similar approaches to extracting features with fully convolutional autoencoders have been used in \cite{abdolghader2021unsupervised, mou2017unsupervised, chen2014deep, zhou2020advances, tao2015unsupervised,ma2016spectral, kemker2017self, ji2017learning, han2017spatial,mei2019unsupervised}, but the subsequent segmentation was not learned in a completely unsupervised setting in these cases.
Although this approach has been shown to be effective at performing generic image classification tasks \cite{li2018discriminatively,guo2017deep} (and has been applied to satellite HSI data \cite{nalepa2020unsupervised}, though only achieving relatively poor segmentation quality), there may be room to further optimise the architecture for processing biomedical HSI data. For Raman and IR HSIs, it is the location of peaks/spectral features along a spectrum that encodes information about molecular contents \cite{shipp2017raman}, and it is the spatial arrangement of different molecules (represented as unique spatial arrangements of peaks or other spectral features specifically found within a subset of each band in the HSI) that can be used to define a tissue type. Therefore, it is important to ensure that the chosen network is optimised to detect spatio-spectral features that only reside in specific regions of the spectral band. However, generic architectures may not be the most effective at achieving this functionality. 

\begin{figure}
    \centering
    \includegraphics[width = \columnwidth]{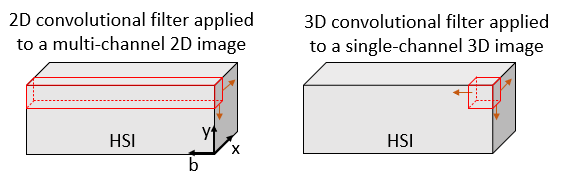}
    \caption{Schematic showing how 2D convolutional filters (red) are applied to multi-channel 2D images, and 3D convolutional filters (also red) are applied to single channel 3D images. 2D filters span the whole spectral band and are `scanned` across the sample in the x-y dimensions \cite{chong2021end,mittal2021survey}. In contrast, 3D filters are typically smaller in the band dimension `b', and take steps in all three dimensions. 2D filters are capable of detecting features that span the whole spectral band (or several features found across the band) but may lack sensitivity to subtle features that may occur only within a specific subset of the band. 3D filters also lack this sensitivity, as they are predisposed to detect more generic features that may occur anywhere within the HSI.}
    \label{fig:2dvs3d}
\end{figure}

In \cite{nalepa2020unsupervised}, HSIs are treated as 3D images composed of single channels and processed with 3D convolutional layers \cite{nalepa2020unsupervised}. However, 3D convolutional layers are not optimised to detect subtle but important features that may be unique to only a portion of the spectral band, instead they are predisposed to learn filters that detect features that may be present anywhere within the whole input volume. As a consequence, they may miss or be much less efficient at detecting subtle but relevant features that reside in specific portions of the spectral band. Networks composed of 2D convolutional layers are typically used to extract features from multi-channel 2D images \cite{lecun2015deep}. However, these are also sub-optimal, as they are designed to detect features that span the whole spectral band, or simultaneously detect multiple features across the whole band \cite{chong2021end,mittal2021survey}. In both cases, the learned representation of patch features may not encode information most relevant to segmentation, spurring a desire for a more efficient feature extraction framework. Fig. \ref{fig:2dvs3d} shows how the filters are applied in both cases, and more details about this problem are provided in Section \ref{sec:auto_enc_details}. 


In contrast to these architectures, the recently proposed UwU-net first processes the input HSI with a series of 2D convolutional layers, ultimately outputting fewer activation maps than the length of the input's spectral band. The resulting activation maps are then split along the channel dimension, where each is fed into its own `inner' U-Net \cite{ronneberger2015u}. Their outputs are then concatenated and processed by another series of convolutional layers to produce the final output. With this architecture, the filters of each inner U-Net are specialised to detect the presence of features encoded in the activation map they are processing. This is unlike generic U-Nets where filters may learn to detect features encoded in all of the activation maps output by the first set of convolutions layers. Experiments have shown that UwU-net architectures outperform more generic U-Nets on supervised image-to-image regression tasks involving HSI data \cite{manifold2021versatile}. In these settings, it is clear that specialising feature extraction in this manner provides a clear advantage over generic 2D architectures. I.e. allowing the network to analyse certain activation maps in isolation at some stages (and also allowing it to learn the kinds of features that should be processed by each inner U-Net) appears to improve the network's ability to detect useful features contained within HSI datasets. Therefore, it is of interest to observe whether this feature extraction framework may confer any advantages for unsupervised segmentation.

An interesting observation is that if the initial convolutional layers used to process the input HSI learn a straightforward compression (e.g. so that each channel encodes features from a given subset of the input HSI's spectral band), then each inner U-Net will learn filters specialised to extract features found within a specific subset of the input's spectral band. This could improve the network's capability to extract features that may reside only in a subset of the input's spectral band, and therefore increase the quality/relevancy of features encoded in the latent representation. With that said, given the architecture is composed of 2D convolutional layers, there is no guarantee that adjacent channels in the activation maps used as inputs to the inner U-Nets will encode information about features residing in similar regions of the input HSI's spectral band. Though unlike more generic architectures, if any of this information happens to be retained, this framework can exploit it. Modifications to the architecture could be made to ensure that the compressed data retains this structure. For example, the downsampling along the spectral band could be performed as a manual pre-processing step with hand-crafted constraints, or by simply averaging neighbouring bands. Though it is expected that allowing the network to learn the most optimal manipulation of its inputs for its architecture (i.e. allowing it to learn which features should be encoded in the inputs to each inner network branch) will deliver superior segmentation quality. Therefore, adapting the original UwU-net framework for end-to-end unsupervised segmentation/clustering remains the focus of this work. 



\subsection{Outline of experiments}
\subsubsection{Synthetic fat/muscle HSIs}
First, we demonstrate how unlike spectral k-means, this spatio-spectral approach enables the segmentation of tissue sections containing similar molecular contents by the way their constituent spectra are arranged in space. This is illustrated using synthetic Raman HSIs of porcine tissue divided into sections, each containing a distinct fat distribution. We compare an end-to-end spatio-spectral clustering strategy and the same approach \emph{not} trained in an end-to-end manner with the results of spectral k-means.
\subsubsection{Real colon HSIs}
To assess our network's ability to segment more complex samples (requiring the extraction of a much larger set of features), we applied it to real IR HSIs of colon tissue. We show that the resultant segmentations have good correspondence with the contents of HE stained adjacent tissue slices used as approximate ground truths. We also compare our results to those acquired with more generic architectures based on 2D and 3D convolutional layers to assess whether any particular strategy may fail to reproduce major morphological features known to be present in the sample.
\subsection{Summary of contributions}
Overall, our main contributions can be summarised as follows:
\begin{enumerate}
    \item We describe the advantages and show the first demonstration of using a fully unsupervised convolutional autoencoder-based end-to-end framework for spatio-spectral clustering/segmentation of biomedical HSI data. Up to now, this approach has only produced relatively poor results on satellite hyperspectral images of geographical landscapes \cite{nalepa2020unsupervised}, or has been used to accurately cluster/group together (but not segment) simple 2D images \cite{li2018discriminatively,guo2017deep} that do not encapsulate the complexity or high dimensionality of biomedical HSIs.
    \item We provide a detailed hypothesis about why generic CAE architectures may be sub-optimal for extracting features for spatio-spectral clustering, and propose a method to improve feature extraction using UwU-net inspired architectures.
    \item We show that our novel architecture appears to produce quality segmentations on real data, indicating its robustness to experimental artefacts and noise in addition to tissue complexity.
    \item We show evidence that the end-to-end clustering approach can help improve segmentation quality compared to workflows with separate compression and clustering steps.
    \item We provide a simulated example that concretely demonstrates how patch-based spatio-spectral clustering enables the segmentation of tissue components by the way their constituent spectra are arranged in space (in contrast to purely spectrum-based segmentation approaches).
\end{enumerate}
\subsection{Code repository}
Some of the code used to generate the results shown here can be found at \url{https://github.com/ciaranbench/unsupervised-HSI-seg}. Much of this is modified code from \cite{guo2017deep} (clustering module/end-to-end framework) and \cite{manifold2021versatile} (UwU-net inspired architecture).

\section{Method: Segmenting synthetic Raman HSIs}
\label{sec:sim_hsi_dec}

\begin{figure}[!htb]
    \centering
    \includegraphics[width=\columnwidth]{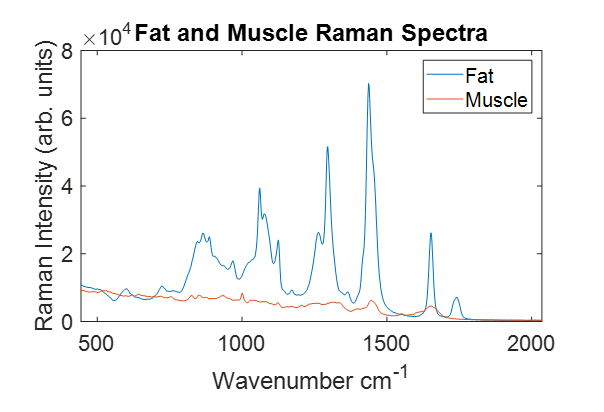}
    \caption{Raman spectrum of porcine fat and muscle samples (back bacon) used to construct synthetic HSIs. These spectra were not pre-processed or normalised, and were acquired with an InVia imaging system (Renishaw, Wotton-under-Edge, UK) with an excitation wavelength of 830 nm, a power of 130 mW, a 50$\times$ long working objective, a 600 L/mm grating, and a 3s$\times$10 exposure time.}
    \label{fig:fat_musc_spec}
\end{figure}
\begin{figure}[!htb]
    \centering
    \includegraphics[width=\columnwidth]{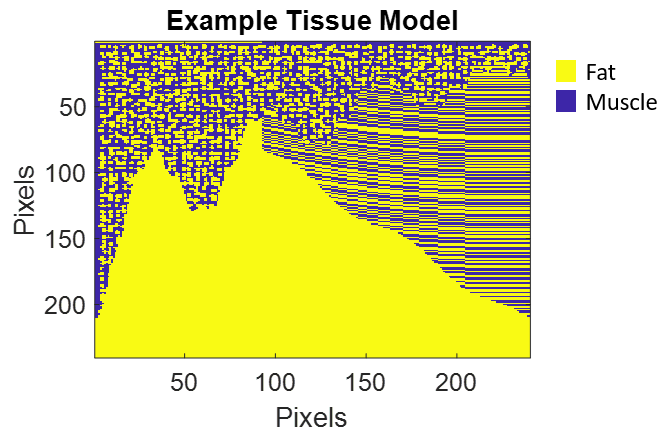}
    \caption{An example $240\times240$ pixel tissue mask used in the construction of a synthetic HSI. Here, blue pixels represent muscle tissue, and yellow pixels represent fat. Three different fat patterns are embedded in the muscle - a solid section, a striped section, and a section containing a random distribution of fat `globules'. The same measured muscle Raman spectrum (acquired from a back bacon sample) is assigned to each muscle pixel. Similarly, the same fat spectrum is assigned to all fat pixels. The resulting HSI has dimensions of $240\times240\times800$.}
    \label{fig:example_muscle_HSI_mask}
\end{figure}

We applied our UwU-net inspired patch-based clustering approach to synthetic Raman HSIs of muscle tissue divided into sections, each containing a distinct fat distribution. We used two training strategies - an end-to-end training framework, and one that performed feature mining and clustering separately. Here we describe the preparation of the training data, details about the architecture used for training, and the two chosen training strategies in depth.
\subsection{Data preparation}
\label{sec:data_prep_sim}
The synthetic HSIs depict generic porcine muscle tissue with fat distributions designed to mimic simple patterns that may be found in real tissue (though only approximately). The muscle and fat spectra used to construct the HSIs (shown in Fig. \ref{fig:fat_musc_spec}) were acquired from back bacon samples (with dimensions of approximately 3 cm$\times$3 cm$\times$3 mm) placed onto stainless steel slides. Prior to measurement, samples were stored at 4-6 $^{\circ}$C. Measurements were acquired with an InVia system (Renishaw, Wotton-under-Edge, UK) with an excitation wavelength of 830 nm, a power of 130 mW, a 50$\times$ long working objective, a 600 L/mm grating, and a 3s $\times$ 10 exposure time. The Raman spectrometer was calibrated using a Silicon peak at 520.5 cm$^{-1}$. Two point measurements were taken at two different positions on the same sample, one containing fat and one containing muscle. The spectra were not pre-processed with a baseline correction or smoothing. 


Each HSI was constructed from a $240\times240$ pixel (referred to as the $x$ and $y$ dimensions respectively) binary tissue mask created by assigning fat patterns to different predefined sections (an example is shown in Fig. \ref{fig:example_muscle_HSI_mask}). Each tissue model contained at least one section of solid fat, one section of striped fat, and one section with randomly distributed `globules' of fat. Each fat pixel was assigned the same fat spectrum. Similarly, one muscle spectrum was assigned to each muscle pixel. 800 Raman shift values were used for each spectrum (ranging from 444.2 cm$^{-1}$ to 2035.2 cm$^{-1}$ in steps of 1.8 cm$^{-1}$). Therefore, each HSI had dimensions of $240\times240\times800$.

Once constructed, each HSI was then decomposed into 3136 $20\times20\times800$ patches, by taking steps of 4 pixels in a raster scan style fashion in the $x$ and $y$ dimensions. All negative values were zeroed. Patches were subsequently normalised by dividing them by the max pixel amplitude from the whole HSI. 

\subsubsection{Alternative methods for synthetic image generation and comments on synthetic data quality}
Arranging measured spectra into custom biomimetic patterns is one of several approaches that can be used to simulate Raman HSI data. For example, Raman spectra can be simulated using models of light propagation in digital tissue models \cite{dumont2021computationally, apra2019simplified}. However, it is challenging to ensure these spectra contain realistic experimental artefacts, limiting our ability to assess how our network may cope with real data.

Despite the fact that our synthetic images are composed of real measured data, our chosen construction method has a few disadvantages. It assumes that our sample i) only contains fat and muscle, ii) all fat pixels and all muscle pixels share the same spectra, and iii) each pixel individually contains only fat or only muscle. In real tissue, none of these assumptions may be completely satisfied. The types of molecules present in a given component may vary from region to region within a sample, producing differences in measured Raman spectra. For example, such changes have been observed in porcine fat spectra acquired at various depths relative to a superficial skin layer, tentatively attributed to the fact that hydrocarbon chains tend to become more ordered (saturated) with depth \cite{lyndgaard2012depth}. Therefore, fat spectra measured from different regions of a pork sample that has been sliced at an angle relative to a skin layer may vary. Physiological parameters like water content (e.g. regions of a tissue slice exposed to air for longer may have less water), and lactic acid \cite{scheier2013measurement} vary subtly within a tissue and will have a corresponding effect on measured Raman spectra. Instrument variability may also cause differences in spectra measured from the same kinds of tissue \cite{pence2013assessing}. Furthermore, in an experimental setting it is possible for a probed region to contain both fat and muscle (i.e. at a muscle/fat interface, or thicker samples), whereas each pixel in our simulated HSIs individually contain only fat or muscle \cite{hedegaard2011spectral,wei2020overview}. Lastly, bone fragments may be present in some cuts of meat, and so there is no guarantee that substances other than fat and muscle will not be present in a given sample.

However, given this is the first reported demonstration of this approach on any kind of biomedical HSI data, we are primarily interested in showing a proof of concept in an idealised scenario. Our simulated HSIs are sufficient for this task, as i) the data shares the same dimensionality of typical biomedical HSIs (thus offering similar challenges with performing efficient feature extraction), ii) their constituent spectra are of biomedical origin (therefore contain at least some of the features we would expect from real tissue), iii) these spectra contain noise and other experimental artefacts (and therefore allow us to demonstrate the capability of our approach to work in the face of some confounding experimental effects), and iv) their segmentation requires the utilisation of spatial features. 

Expanding on point iv), the discrepancy between a purely spectrum-based segmentation approach and one that utilises both spatial and spectral features may not be as drastic in real tissue as may be suggested by the experiments on our synthetic images. Different tissue regions containing the same kinds of molecules may be differentiated by unique spectral features alone due to experimental factors. For example, consider a hypothetical tissue with two regions $A$, and $B$. Let us assume that both contain only fat and muscle. Region $A$ has fat and muscle densely packed at a fine scale (e.g. smaller than the area typically illuminated by the Raman probe when used to measure a spectrum), whereas region $B$ mostly consists of muscle with large globules of fat dispersed at a scale much larger than the illuminated region. Because each probed region in $A$ will contain both fat and muscle, the resultant spectra will contain contributions from both. In contrast, most of the spectra acquired from $B$ will be acquired from `pure' fat or muscle. As a consequence, spectra measured from region $A$ will be distinct from those measured in $B$ - therefore, spectral information alone may be used to segment $A$. With that said, this represents a hypothetical scenario, and the likelihood of encountering this effect is unclear. Nevertheless, it illustrates how in practice purely spectrum-based approaches may be more effective at differentiating tissue regions than our simulation study suggests. Some of these factors may be effectively replicated with a real physical phantom and should be the subject of future work.

\subsection{Training on a single image}
In typical deep learning applications, the aim is to train a network so that it generalises well to unseen data (requiring no additional training or modifications to its parameters). Therefore, evaluations of network performance are typically performed by processing unseen test data using a pretrained model. However, we are not interested in finding some pretrained model that is applicable to all images, but rather, a general algorithmic framework that can produce a model that accurately segments any singular image it may receive as an input (akin to how spectral k-means can be applied individually to each image in a given dataset). Therefore, a network's input will consist of patches acquired from a single HSI. Furthermore, the algorithm is run independently and from scratch for each HSI it may receive. The effectiveness of our algorithm is assessed by observing the accuracy of the resultant segmentation produced from this singular HSI.

With that said, overfitting is still something that should be avoided as this will reduce the quality of the compressed representation, and hence, negatively impact the quality of the resultant segmentation. Therefore, we still use a validation loss to assess the degree of overfitting and hence the usefulness of the feature encoding even if we are not aiming to make the network generalisable.

Even though a single image training approach is used here, it would be possible to train a CAE that generalises well to different tissue samples. This could be trained by feeding it a large set of images. After a lengthy training phase, this would ultimately speed up the time needed to acquire segmentations. However, an advantage of our single-image training approach is that the full expressive power of the network is being used to learn the compression and segmentation for the single example. Segmentations produced with a network trained on a large set may be comparable, but will ultimately suffer from some generalisation error. 

\subsection{Network architecture outline}
\label{sec:arch_summary}
The architecture features two modules, an autoencoder module $A$ and a clustering module $C$ (see Fig. \ref{fig:net_arch}).
\subsubsection{Autoencoder module}
Here, the autoencoder module is used to `mine' features from the input patches that can be used for their subsequent clustering. A convolutional autoencoder (CAE) learns to i) detect and compress spatio-spectral features contained in each image patch into a set of low-dimensional latent vectors, and ii) subsequently reconstruct the input from this compressed representation. The latent vectors encode information about the spatial and spectral features contained in the input. As a consequence, they can be used to group together patches containing similar contents.

More specifically, the autoencoder module $A$ consists of two components: i) an encoder $A_E$ that learns a mapping from an input HSI patch $x_i \in X$ to a set of low-dimensional latent vectors $l_{i,m} \in L$, $A_E: X \rightarrow L$ were $i$ indexes each patch, and $m$ indexes the latent vector produced by each CAE branch (explained in Section \ref{sec:auto_enc_details}) and ii) a decoder $A_D$ that learns a mapping from the input patch's set of latent vectors $l_{i,m}$ to a reconstruction of the original input $\hat{x_i} \in \hat{X}$, $A_D: L \rightarrow \hat{X}$. Further details about the structure of $A$ and the motivation for its design are discussed in Section \ref{sec:auto_enc_details}. 
\subsubsection{CAE+k-means segmentation}
Once trained, the set of latent vectors associated with each input image patch will encode information that enables the decoder to perform the reconstruction. This should consist of information about the spatial and spectral features contained in each patch. Regions of the HSI that share similar spatial and spectral features can be segmented by grouping them with patches that have similar latent vectors. This can be performed by i) concatenating the latent vectors produced from each patch into a single 1D vector ($\overline{l_i}$), then ii) clustering them with the corresponding $\overline{l_i}$ produced from all other patches (e.g. with k-means), and then iii) performing a reconstruction step to acquire the resultant segmentation image. Taking inspiration from \cite{guo2017deep}, we refer to this style of segmentation as \textbf{`CAE+k-means'} clustering (see section \ref{sec:CAE+kmeans_description}). However, with this approach, the CAE does not compress input patches in a way that is optimised for their subsequent clustering, instead, solely prioritising optimal reconstruction quality. 
\subsubsection{Clustering module and end-to-end clustering}
To improve the quality of cluster assignments, the architecture also contains a clustering module $C$ that learns a mapping $C: L \rightarrow O$, where $o_{i,j} \in O$ is a set of `soft-assignments', each describing the confidence that the patch (indexed with $i$) should be assigned to each of the clusters (indexed with $j$, where the total number of clusters is a user-defined parameter). The clustering loss is used with the reconstruction loss to update the parameters of $A$ and $C$ to ensure that features are extracted and compressed in a way that optimises the `accuracy' of their subsequent cluster assignment performed by $C$. In $C$, the set of latent vectors associated with each input patch are concatenated into a 1D vector $\overline{l_i}$ that is then processed by a clustering layer, where the Student's t-distribution is used to compute $o_{i,j}$ for each $\overline{l_i}$. We call this style of clustering \textbf{`end-to-end clustering'}. Details about how the clustering loss is calculated and utilised in the training process are described in Section \ref{sec:end-to-end_cluster_description}. 
\begin{figure*}
    \centering
     \includegraphics[width=1.95\columnwidth]{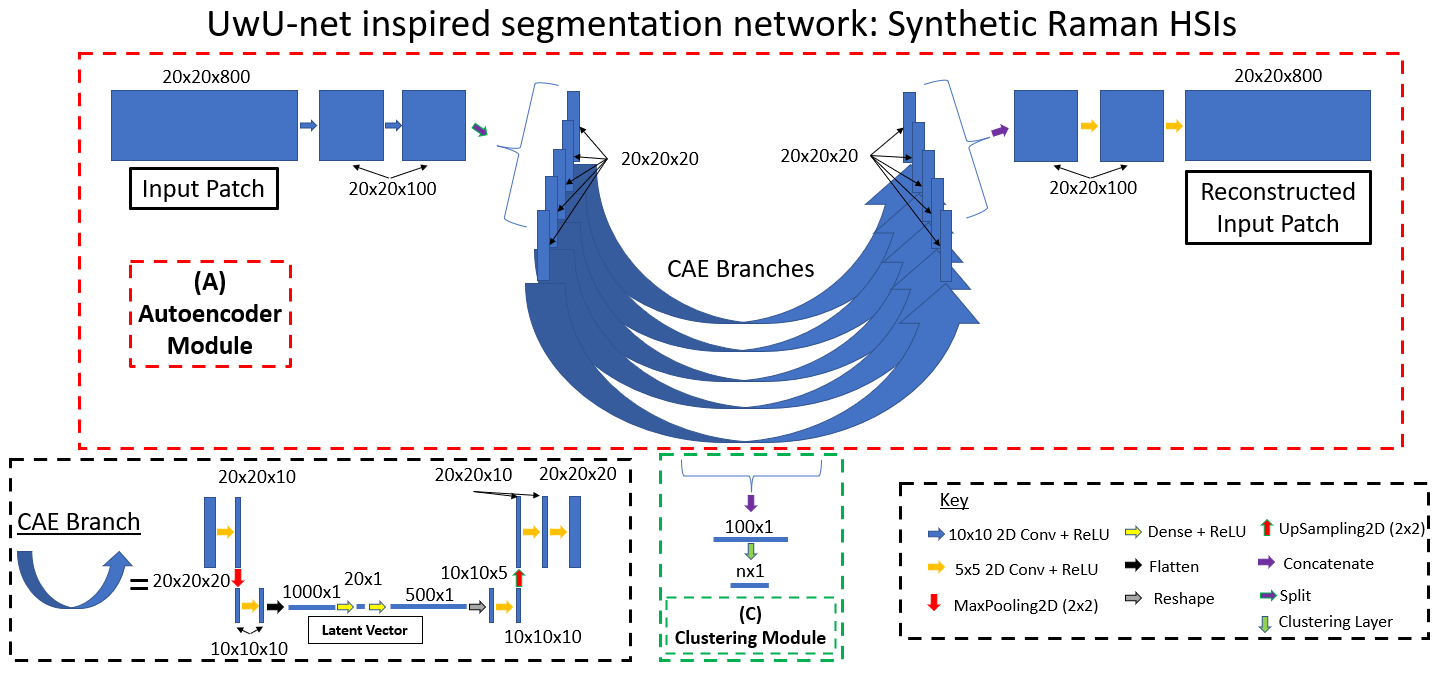}
    \caption{UwU-net inspired segmentation network architecture. The convolutional autoencoder module (A) is used to mine features from the input HSI patch, and compress them into a 100 element latent vector. The architecture is inspired by the UwU-net that is hypothesised to provide a more effective feature extraction framework compared to more generic CAE architectures. The clustering module (C) outputs a probability that the patch belongs to each cluster. With the end-to-end training scheme (A) is initially trained in isolation. After this pretraining step, both (A) and (C) are trained together.}
    \label{fig:net_arch}
\end{figure*}
\subsubsection{Discussion of architecture choice and further network details}
\label{sec:auto_enc_details}
Here, we discuss the motivation for using the UwU-net inspired architecture and describe the structure of the autoencoder and clustering modules in greater detail. 

Information about molecular contents is given by the presence of peaks/spectral features lying within a certain range of the spectral band of the IR or Raman spectra. Thus, it is expected that many of the features relevant to tissue classification will reside in a localised subset of the spectral band. Ideally, the autoencoder module should be optimised to extract these kinds of features to improve efficiency and to increase the chances that all relevant features (even those that may be subtle) can be detected. 

Autoencoders processing multi-channel 2D images are usually composed of a series of 2D convolutional layers and pooling layers. However, each filter (dimensions $n  \times n \times b$) is predisposed to detect features that may span the whole spectral band, or multiple features throughout the band (decreasing sensitivity to subtle features that reside within a subset of the band). E.g. a filter may learn to detect a subtle feature that occurs at one point in a spectrum alongside a large set of other spectral features. However, the contribution to the activation from this subtle feature could be quite low. Even though a filter may have some sensitivity to it, the resultant activation will not change much if it is present or not if the rest of the features found within the band that the filter detects are also present. With that said, it is possible that the layer may learn filters that only detect features that primarily occupy a specific portion of the spectral band (e.g. having weight values of/near zero everywhere except the region of the band that is of interest), but this is not guaranteed and is likely less efficient than using a more optimised framework for feature extraction. As a consequence, the compressed representation may not robustly encode information about whether subtle spatio-spectral features unique to a subset of the band that are relevant to segmentation are present in a patch.


In other cases, HSI patches have been treated as single channel 3D images and processed by autoencoders composed of 3D convolutional layers \cite{li2017spectral,wang2019classification,he2017multi}. However, these are also sub-optimal. The $n \times n \times n$ filters will learn to detect features spanning the filter's volume that may reside anywhere within the input patch, and are not optimised to detect features that may specifically occur in a subset of the spectral band. Though at least here, the resultant activation map will have information about where in the input the detected feature resides in, retaining context about its position along the spectral band (unlike 2D CNNs). With that said, there are no constraints preventing it from mostly detecting generic features that commonly occur anywhere in the volume, potentially missing or being less effective at detecting subtle features that reside within a certain subset of the spectral band. Additionally, 3D convolutional layers often require more processing power and memory to train compared to 2D CNNs, making them generally less appealing to work with. 

In contrast to these approaches, the recently proposed UwU-net first processes the input HSI with a series of 2D convolutional layers outputting fewer activation maps than the length of the input's spectral band. The resulting activation maps are then split along the channel dimension, where each is fed into its own `inner' U-Net \cite{ronneberger2015u}. The activation maps produced by the final layer of each inner U-Net are then concatenated together, and processed by another series of convolutional layers to produce the desired output. Therefore, instead of having filters that may learn to detect hierarchical features composed of features encoded in all of the activation maps produced by the first set of convolutional layers (as would be the case with a generic U-Net), the filters of each inner U-Net are specialised to detect features unique to the activation map used as their input. Empirical evidence has shown that allowing the network to analyse some activation maps in isolation as opposed to all at once (and also allowing it to learn the best way to collate features for further processing by each inner U-Net) provides a clear advantage over generic 2D architectures for image-to-image regression tasks \cite{manifold2021versatile}. Therefore, it is of interest to observe whether this feature extraction framework may provide any advantages for unsupervised segmentation.

It is interesting to note that if the first few convolutional layers learn to encode features from various subsets of the input HSI's spectral band into distinct channels (i.e. they learn a straightforward compression along the spectral dimension), then each inner U-Net will learn filters specialised to extract features residing in a specific subset of input HSI's spectral band. This could increase the quality/relevancy of features encoded in the latent representation. Though, given the input patch is initially processed with 2D convolutional layers, there is no guarantee that adjacent channels in the activation maps will encode information about features residing in similar regions of the input patch's spectral band. But, as stated previously, if any of this information is retained, this framework can exploit it. In any case, the architecture could be modified to help preserve this structure in the data, for example, by performing the downsampling along the spectral dimension in a manual pre-processing step with hand-crafted constraints. Though allowing the network to learn the compression in an end-to-end framework (i.e. allowing it to learn which features should be encoded in the inputs to each inner network branch) will ensure that this downsampling is optimised for processing by the rest of the architecture, ultimately improving segmentation quality. Therefore, in this work we consider the fully learned approach.

We adapt this architecture for an unsupervised patch-based clustering approach (see Fig. \ref{fig:net_arch}). Instead of whole HSIs, the network input is a ($20\times20\times800$) patch of the HSI. The input HSI patch is compressed along the spectral dimension to $20\times20\times100$ as a first step towards reducing the number of parameters required to process/extract spatio-spectral features from the input. These are then split into 5 equal components along the channel dimension (e.g. $k_1$, $k_2$, $k_3$, $k_4$, and $k_5$, though any desired number of divisions can be chosen in principle).  Each $k$ is then used as the input to its own `inner' convolutional autoencoder (CAE). The outputs of these parallel inner CAEs are then concatenated together, and processed by a series of 2D convolutional and upsampling layers to produce the desired output. All convolutional layers have a stride of 1.

There are two differences between our architecture and that used in the original UwU-net paper. Firstly, the `inner' parallel U-Nets have been replaced with CAEs (i.e. essentially U-Nets \emph{without} skip connections). The skip connections are removed to ensure that the network learns how to detect and compress information about patch-defining features into latent vectors. This is not guaranteed with an encoder-decoder style architecture \emph{with} skip connections, as information can propagate through the architecture without passing through the contracting path that forces the network to learn a compressed representation. Secondly, in the original UwU-net paper, the inputs to each inner U-Net is a single activation map. Here, we perform a less aggressive compression, where the input to each inner CAE is a multi-channel 2D activation map. This is done to increase the amount of information available to each inner CAE.



The second module, $C$ (inspired by the clustering module used in \cite{guo2017deep,xie2016unsupervised,nalepa2020unsupervised}) takes the latent vectors from each parallel CAE ($l_{i,1}$, $l_{i,2}$, $l_{i,3}$, $l_{i,4}$, $l_{i,5}$, where $i$ denotes the index of the HSI patch being processed) as inputs, first concatenating them into a single vector $\overline{l_i}$, and then processing this with a clustering layer. The resultant output is the probability of the patch belonging to each cluster. 

\subsection{Training schemes}
Here we provide further details about how each training scheme mentioned in the outline in Section \ref{sec:arch_summary} is executed. 
\subsubsection{Training details - CAE+k-means}
\label{sec:CAE+kmeans_description}
In this approach, the architecture consists only of $A$ that is trained to compress the contents of each patch into latent vectors and reconstruct each patch from these compressed representations. $A$ was trained with the mean square error as the loss function with a batch size of 64, a learning rate of 0.001, and Adam as the optimiser. The HSI of another tissue sample was used as a validation set to monitor the degree of overfitting, ensuring $A$ learned to compress useful features. Training was terminated with an early stopping approach. Once $A$ was trained, the latent vectors for each patch (indexed with $i$) $l_{i,1}$, $l_{i,2}$, $l_{i,3}$, $l_{i,4}$, $l_{i,5}$ were concatenated into a single 1D vector $\overline{l_i}$. The $\overline{l_i}$ for all patches were then clustered using the k-means++ algorithm in MATLAB R2021b with default parameters, and with 3 clusters. With this approach, the compression and clustering steps are completely separate, therefore, the compression/feature detection learned by $A$ is not optimised for the subsequent clustering step. The pretraining loss curves for most of the networks mentioned in this work are shown in Appendix \ref{sec:loss_curves}.

\subsubsection{Training details - End-to-end clustering}
\label{sec:end-to-end_cluster_description}
With the end-to-end clustering approach the network consists of both $A$ and $C$. $A$ is initially trained in isolation using the same procedure as \ref{sec:CAE+kmeans_description}. Then, the cluster centres in $C$ are initialised by i) acquiring the latent vector for each patch ($\overline{l_i}$) in the training set with $A$ and then ii) applying the k-means++ algorithm (sklearn) to cluster these latent vectors. Once initialised, all modules are trained together with a weighted loss function (explained shortly). It is suggested in \cite{guo2017deep} that the unregularised/unconstrained inclusion of the clustering loss may `distort' the latent representation (i.e. reduce how effectively a patch's spatio-spectral features are encoded in their respective latent vector), negatively impacting the cluster assignment accuracy. Continuing to optimise $A$ during the combined training stage helps ensure that the encoded features can still be used to accurately reconstruct the input, and therefore, robustly encode information about the spatio-spectral features present in each patch. Using autoencoders to strengthen feature representation has also been utilised in \cite{peng2016deep,goodfellow2016deep}. Pretraining $A$ in isolation ensures that the initialised cluster centres are based on robust representations of patch features that have not suffered from this distortion.

Latent vectors are clustered using the Student's t-distribution, which outputs `soft' cluster assignments \cite{guo2017deep,xie2016unsupervised}:
\begin{equation}
    o_{i,j} = \frac{(1+\| \overline{l_i} - u_j \|^{2})^{-1}}{\sum_{j}(1+\|\overline{l_i} - u_j \|^2)^{-1}},
\end{equation}
where $u_j$ are the cluster centres (trainable parameters), $\overline{l_i}$ is the latent vector, and $o_{i,j}$ can be interpreted as the probability that the latent vector $\overline{l_i}$ belongs to cluster $u_j$.

The network is (in part) trained to reduce the discrepancy between the output `soft' cluster assignments ($o_{i,j} \in O$ where the patch is indexed with $i$ and each cluster with $j$), and the target soft cluster assignments for each patch ($p_{i,j} \in P$). The target assignments $p_{i,j}$ are computed using, 
\begin{equation}
    p_{i,j} = \frac{o_{i,j}^2/f_{j}}{\sum_{j}(o_{i,j}^2/f_{j})},
\end{equation}
where $j$ is used to index the cluster centres, and $f_j = \sum_{i} o_{i,j}$ are the cluster frequencies (number of examples assigned to each cluster). These target assignments strengthen predictions/improve cluster purity, place greater emphasis on cluster labels assigned with high confidence, and prevent large clusters from dominating by incorporating a normalisation \cite{xie2016unsupervised,guo2017deep}. In essence, we leverage an approach similar to self-training where confident assignments are assumed to be more accurate and are used to `guide' the assignments of patches with similar content.

It is important to note that the use of this clustering module may only improve cluster assignments if the more confident predictions produced with k-means during the initialisation step are accurate. Based on the results acquired with the CAE+k-means approach (e.g. see Fig. \ref{fig:res_1}), we believe this assumption to be accurate for our dataset. The Kullback-Leibler divergence (KL) is used to determine the degree of similarity between the target assignments $P$ and output assignments $O$ and makes up part of the objective function in the end-to-end clustering approach:
\begin{equation}
    L_{Clust}= \mathrm{KL}(P||O) = \sum_i\sum_j p_{i,j} \log{\frac{p_{i,j}}{o_{i,j}}}.
\end{equation}

Two outputs are produced via the forward pass: the cluster assignment for the patch, and a `copy' of the patch reconstructed from its compressed representation. The model is trained to minimise $V$, the weighted sum of the resultant reconstruction loss and clustering loss: 
\begin{equation}
    V = \gamma L_{Clust} + L_{Recon}, 
\end{equation}
where $\gamma=0.1$ is a hyperparameter that weights the contribution of the clustering loss to $V$, and $L_{Recon}$ is the mean squared error.

The target soft-assignments for each patch are updated after each epoch. Training is terminated if one of two conditions were met: either i) once the total number of changed cluster assignments (assessed at the end of each epoch) was less than .1\%, or ii) after 10 epochs had elapsed. The latter threshold was chosen to allow for repeat clustering sessions to be completed in only a few hours if necessary. These may be required as the resultant clusters are sensitive to the k-means initialisation, which itself is sensitive to the randomly initialised cluster positions. The network was pretrained with a batch size of 64, a learning rate of 0.001, and with Adam as the optimiser. The CAE pretraining was conducted using the same procedure described in Section \ref{sec:CAE+kmeans_description}.

\subsection{Processing outputs: reconstruction of segmentation images}
The procedure for reconstructing each segmentation image is similar to that implemented in \cite{nalepa2020unsupervised}. The network is used to assign a particular cluster group (i.e. the cluster group with the highest probability) to each input HSI patch, referred to here as the patch's cluster ID. A `cluster patch' is produced for each input HSI patch - this is a 2D matrix with the same $x$ and $y$ dimensions as the HSI patch where each element contains the value of the patch's assigned cluster ID. The resulting segmentation map is reconstructed by tiling these cluster patches in an overlapping fashion, placing them in the same location in the x-y dimensions as their corresponding HSI patch. This tiling starts with the HSI patch at the top left corner, moving right along the x-dimension to the end of the row, and then starting again from the left most patch of the next row down and so on until all patches have been exhausted. The corresponding segmentation image will have the same $x$ and $y$ dimensions as the whole HSI. 

\section{Method: Segmenting real IR-HSIs}
\subsection{Preparation of real IR-HSIs of colon tissue}
We applied end-to-end clustering (using the procedure described in Section \ref{sec:end-to-end_cluster_description}) on a set of three colon HSIs from the Minerva dataset (\url{http://minerva-project.eu/}) \cite{fotonik2015minerva}. Each image has a corresponding HE slide of an adjacent tissue slice. The three images shown here were chosen based on two criteria: i) they had a large number of glands present (therefore allow us to assess the network's ability to segment a large set of components), and ii) their corresponding HE slide shows good correspondence with tissue contents as determined with spectral k-means (i.e. their HE slide can be used as an approximate ground truth).

Colon slices contain several tissue types and components. The most prominent being the intestinal glands, which themselves are composed of several subcomponents: the lumen, epithelial cells, stroma, and nuclei \cite{rathore2019segmentation}. Given the pixel size of the images ($5.5\times5.5$ $\mu m^{2}$), it is not clear whether these subcomponents may be resolvable. At the very least, we aim to segment the areas occupied by glands, as well as other neighbouring tissue types that can be clearly observed in the corresponding HE slides.  

The HE slide corresponding to each HSI depicts the morphology of an adjacent tissue slice. Therefore, the presence and shape of some components may not exactly reflect those found in the tissue sample depicted by the HSI. With that said, the contents are expected to be similar, e.g. glands should remain clustered in similar regions. Though in some cases, the shapes and number of glands can differ significantly. Furthermore, the HE staining does not reveal all regions that may have different molecular contents, instead highlighting nuclei and the extracellular matrix/cytoplasm to reveal differences in morphology \cite{chan2014wonderful}. Ultimately, the HE slides allow us to evaluate whether any architecture fails to reproduce major morphological features of the samples (i.e. whether they generally appear in their expected locations and with similar shapes). Though, it is evident that these slides can not be used to precisely evaluate the quality of the segmentations produced with any of the chosen architectures.

The images were acquired with an Agilent 620 FTIR microscope coupled with an Agilent 670 FTIR spectrometer with a Globar® light source, and a liquid-nitrogen cooled 128 × 128 FPA detector. The resultant images have a $5.5 \times 5.5 \mu m^2$ pixel size with a $704 \times 704 \mu m^2$ field of view. Measurements were acquired in the mid-IR spectral range of 1000–3800 cm$^-1$ at a spectral resolution of 4 cm$^{-1}$. Samples were ‘electronically de-paraffinised’ using a modified extended multiplicative signal correction. Further information about sample preparation, image preprocessing, and instrument details can be found in \cite{nallala2016high}.

Images were prepared by taking $10\times10$ patches in steps of 2 pixels using the same procedure described in \ref{sec:sim_hsi_dec}.
\subsection{Learning unsupervised segmentation with three different CAE-based architectures}
\label{sec:real_HSI_method}
Three different architectures were applied to the IR HSIs (all sharing the same two-component structure of the architecture shown in Fig. \ref{fig:net_arch}, and trained in an end-to-end manner): one where $C$ was a generic 3D CAE (Fig. \ref{fig:3dcnn}), one where $C$ was a generic 2D CAE (Fig. \ref{fig:2dcnn}), and a UwU-net inspired architecture (Fig. \ref{fig:uwu_minerva})). An early stopping strategy was used to determine the stopping point of the CAE pretraining step for the UwU-net and the generic 2D CAE. Each network was trained with Adam as the optimiser, a batch size of 58, and a learning rate of 0.001. Once pretrained, both modules were trained together until less than 0.1\% of assignments changed after a given epoch, or after 50 epochs.

The generic 3D CAE architecture was pretrained for 40 epochs, and the maximum number of epochs for the combined training step was set to 20 to limit the total training time to approximately one day. In some cases, the training of the generic 2D and 3D CAEs had to be initiated a few times before the expected exponential decay of the validation loss was observed. 

In an attempt to give both styles of 2D networks similar expressive power, the generic 2D CAE network was constructed to utilise a similar number of filters and the same size latent vector as the UwU-net inspired architecture. This was done to ensure any observed differences in their performance could be attributed to differences in the splitting/branching of the layers. The 3D network was coded to have a similar structure as the 2D networks, but featured fewer layers and filters to help reduce computational time. It is not clear how much this may have affected the expressive power of this network relative to the 2D networks, as the architecture is manipulating information about features detected with a 3D convolutional filter as opposed to those detected with a multi-channel 2D filter (i.e. perhaps fewer 3D filters are needed to achieve similar expressive power as the 2D networks). Therefore, it is important to bear in mind that the comparison shown here is not necessarily between two networks with similar expressive power relative to their dimensionality. Nevertheless, the comparison still has some value in the sense that it shows the results that can be produced with either architecture given a clinically relevant time window of 1 day for training (the time between image acquisition and analysis of morphological information should ideally be as short as possible). These network architectures are shown in Appendix \ref{sec:generic_cae}

\section{Results}
\subsection{Segmentation of synthetic Raman HSIs}
\begin{figure}[!htb]
    \centering
    \includegraphics[width=.9\columnwidth]{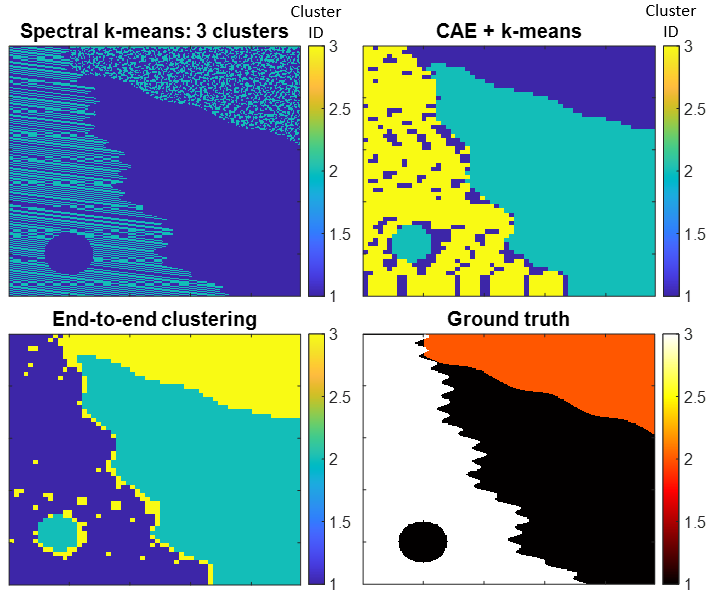}
    \caption{All images shown here have been assigned random colour scales. Top left: results of applying spectral k-means with 3 clusters to a synthetic HSI (Sample 1). Here, fat and muscle pixels are distinguished accurately, but this method can not segment each region containing a unique fat pattern. This image shows which pattern occupies each area of the tissue. Top right: segmentation produced with CAE+k-means. Bottom left: segmentation produced with the end-to-end clustering approach that has significantly improved the quality of the segmentation in the striped region. Bottom row: an image of the tissue mask used to generate the synthetic HSI, showing the extent of each distinct region.}
    \label{fig:res_1}
\end{figure}
\begin{figure}[!htb]
    \centering
    \includegraphics[width=.9\columnwidth]{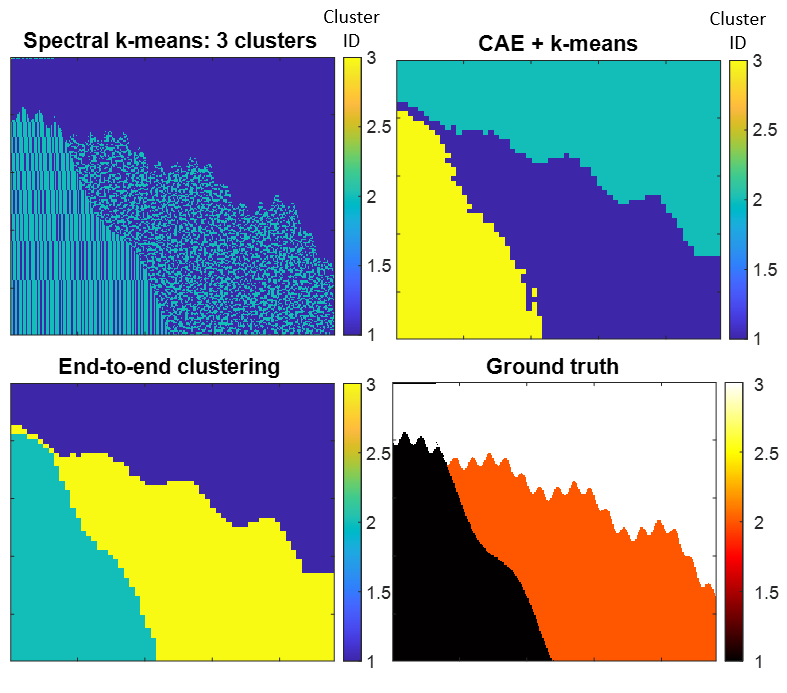}
    \caption{Output segmentations and ground truth for Sample 2 (synthetic HSI).}
    \label{fig:res_2}
\end{figure}
\begin{figure}[!htb]
    \centering
    \includegraphics[width=.9\columnwidth]{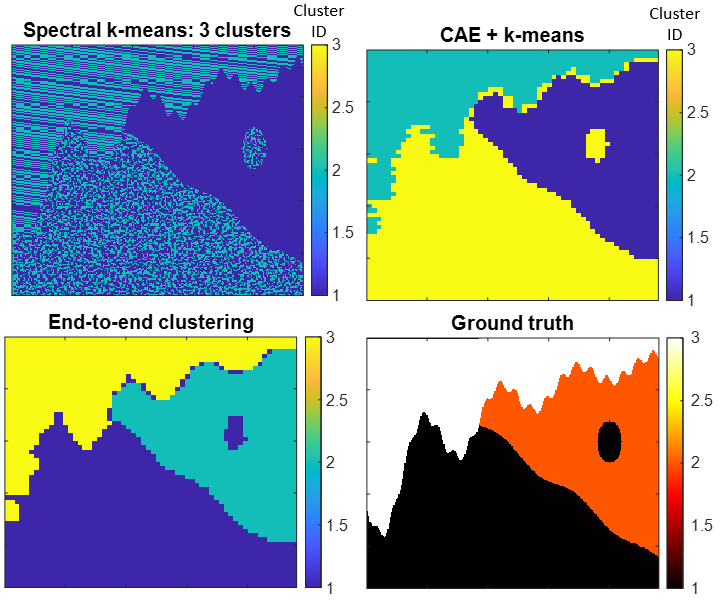}
    \caption{Output segmentations and ground truth for Sample 3 (synthetic HSI).}
    \label{fig:res_3}
\end{figure}

As expected (see Figs. \ref{fig:res_1}, \ref{fig:res_2}, and \ref{fig:res_3}), spectral k-means was able to accurately differentiate muscle and fat pixels (a trivial task, given the properties of each image). However, this approach can not be used to segment each tissue section as it does not utilise information about their constituent spatial features. These figures also show the results of our segmentation algorithms (CAE+k-means and end-to-end clustering) that utilise both spatial and spectral features. These were successful in distinguishing each tissue section. A normalised mutual information (NMI) score and adjusted Rand score (ARS) were used to compare the accuracy of the CAE+k-means and end-to-end clustering approaches (as can be seen in Tables \ref{tab:NMI_sim} and \ref{tab:AR_sim}). For Samples 1 and 3, the end-to-end clustering scores were notably higher than the CAE+k-means, and for the remaining sample, the accuracy of both techniques were comparable. The benefits of the end-to-end approach are most clearly observed in Sample 1 in Fig. \ref{fig:res_1}. Here, several regions within the striped section of the tissue are erroneously assigned to the same class as the globule section of the tissue using the CAE+k-means approach. The number of erroneous cluster assignments is significantly reduced in the end-to-end segmentation output. All segmentations are offset from the ground truth by a distance related to the tiling step size that lowers the segmentation quality scores. This is discussed further in Section \ref{sec:seg_real_hsis}.

\begin{table}[]

 \caption{Segmentation accuracy (NMI Score): synthetic HSIs}
 \begin{center}
\begin{tabular}{lllll}
 &  &  &  &  \\
Example \# & NMI (CAE+k-means) & NMI (End-to-end) &  &  \\
1          & 0.59              & 0.68             &  &  \\
2          & 0.80              & 0.81              &  &  \\
3          & 0.66              & 0.71             &  & 
\end{tabular}
\label{tab:NMI_sim}
\end{center}
\caption{Segmentation accuracy (ARS): synthetic HSIs }
 \begin{center}

\begin{tabular}{lllll}
 &  &  &  &  \\
Example \# & ARS (CAE+k-means) & ARS (End-to-end) &  &  \\
1          & 0.66              & 0.76             &  &  \\
2          & 0.85              & 0.86             &  &  \\
3          & 0.72              & 0.78             &  & 
\end{tabular}
\label{tab:AR_sim}
 \end{center}
\end{table}


For all samples, patches found near the boundary between two tissue regions were clustered into the same group containing fat globule patches (i.e. their latent vectors were most similar to those representing globule patches). The solid and striped sections have distinct features that are fairly consistent across all patches, whereas there is greater variability and randomness in the features found in globule patches. Therefore it is not entirely surprising that patches with an assortment of different fat patterns adjacent to one another would be grouped into the `globule' class. The extent of this decreased slightly with the end-to-end clustering approach. A smaller patch size might help reduce the spatial extent of these artefacts, though this should be done with caution as each tissue region is defined by spatial patterns that span a certain range, and the spatial context represented in each patch should ideally span this scale.

It is important to note that these results do not indicate how successful this approach may be in more practical settings as our model tissues represent idealised samples for several reasons. These are described in depth in Section \ref{sec:data_prep_sim}. Nonetheless, this experiment demonstrates our UwU-net inspired architecture's capability to utilise both spatial and spectral information to segment tissue regions primarily differentiated by the way their constituent molecules are arranged in space, and that training it in an end-to-end matter can improve the accuracy of the resultant segmentation. 

There are two other limitations that are important to note. The first is that this algorithm may only be used to segment regions whose defining features span the patch size. Though patches of various dimensions can be easily accommodated by altering the network architecture. Secondly, the accuracy is dependent on the initialisation of cluster centres performed with k-means. If there is significant class imbalance (i.e. one tissue class occupies a smaller area than the others) then this could lead to errors in the resultant segmentation \cite{franti2018k}.

\subsection{Segmentation of real IR colon HSIs}
\label{sec:seg_real_hsis}

All three CAE architectures produced segmentations qualitatively comparable to the corresponding HE stains. Interestingly, all contained prominent artefacts at the interface between the tissue sample and background - different cluster groups appear to layer on top of each other to form the boundary. We hypothesise that unique latent vectors are required to reconstruct patches containing varying proportions of background and tissue, resulting in the border being defined by a large number of cluster groups and consequently the production of these `layer-like' artefacts. Despite the presence of these artefacts, the border regions appear to maintain the same morphology as that depicted in the HE stain - the boundaries remain smooth, and the shape of the sample remains easy to assess. Though, the various border classes do not appear to correspond with distinct morphological features, and therefore we advise to disregard them when assessing tissue contents.

Given the HE slides depict an adjacent tissue slice and thus only provide approximate information about the morphology of the sample's components, a precise assessment of segmentation quality (or comparisons between all three architectures) is not possible. Though, it is expected that the morphology of the adjacent slice should share similar structural characteristics as the sample depicted in the HSI (e.g. glands gathered in similar locations). Therefore, the HE stains allow us to assess whether any particular architecture may completely fail to segment any major morphological features expected to be present in the sample. More specifically, a broad measure of segmentation quality may assessed by  whether they contain most of the components found in their corresponding HE slide, and whether these components reside in similar locations.

\begin{figure}
    \centering
    \includegraphics[width=.95\columnwidth]{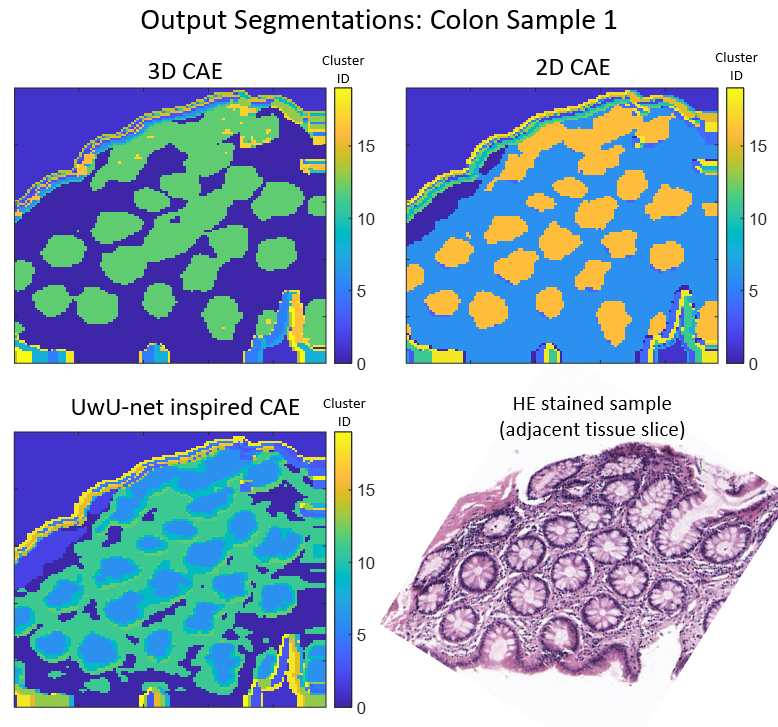}
    \caption{Segmentations of an IR-HSI of a real human colon sample produced with end-to-end clustering using a 3D CAE based architecure (top left) (architecture shown in Fig. \ref{fig:3dcnn}), a 2D CAE based architecture (architecture shown in Fig. \ref{fig:2dcnn}), and an UwU-net inspired architecture (bottom left) (architecture shown in Fig. \ref{fig:uwu_minerva}). The HE stained adjacent tissue slice used as an approximate ground truth is depicted at the bottom right. All three architectures produced comparable results, and appear to be able to segment the locations of most glands. Though, a precise evaluation of segmentation accuracy is not possible given the approximate nature of the ground truth.}
    \label{fig:minerv_seg_1}
\end{figure}

\begin{figure}
    \centering
    \includegraphics[width=.95\columnwidth]{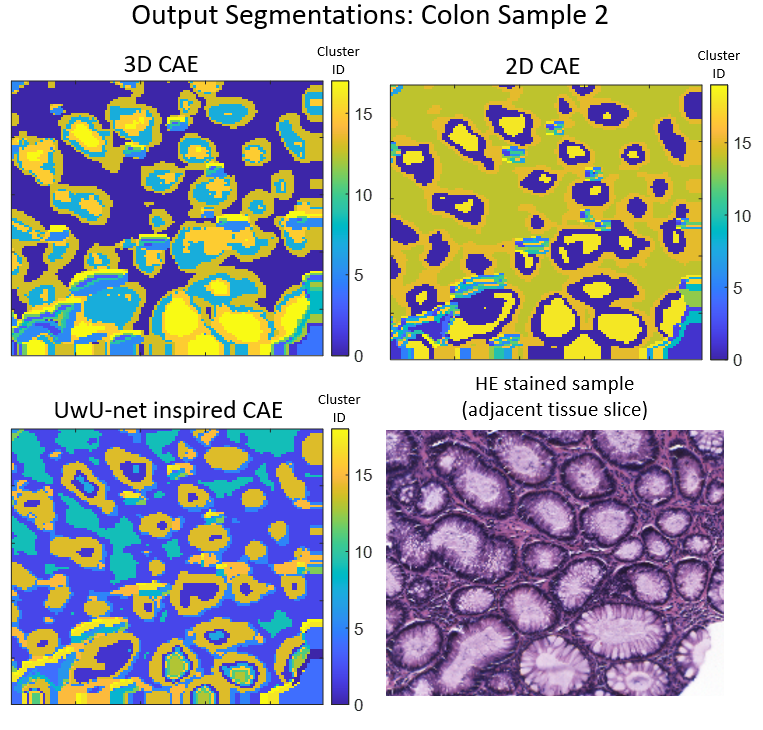}
    \caption{Output segmentations and HE slide for colon Sample 2.}
    \label{fig:minerv_seg_2}
\end{figure}

\begin{figure}
    \centering
    \includegraphics[width=.95\columnwidth]{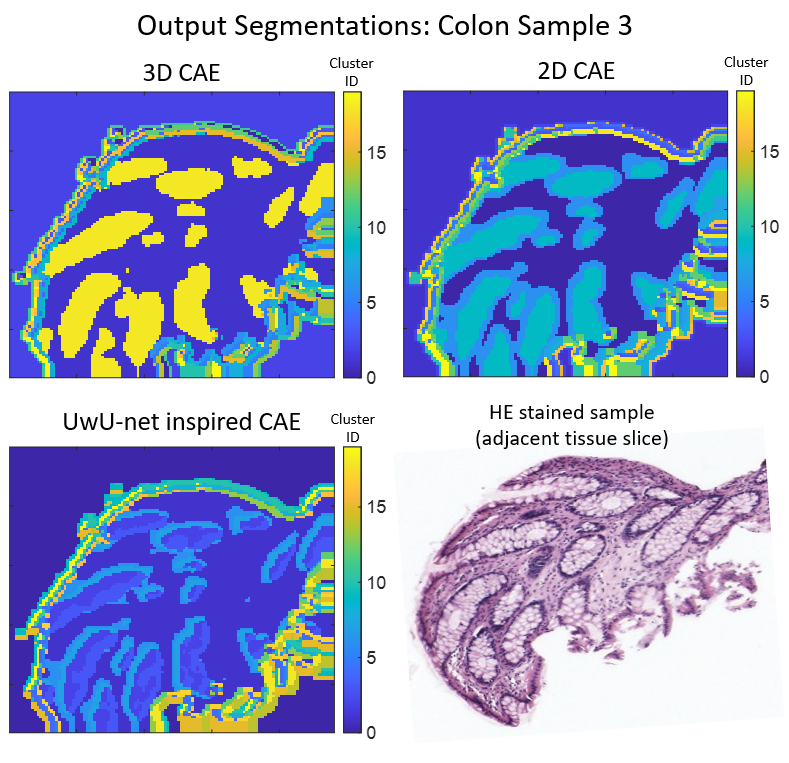}
    \caption{Output segmentations and HE slide for colon Sample 3.}
    \label{fig:minerv_seg_3}
\end{figure}

Each architecture was capable of producing segmentations that satisfy this criteria, demonstrating each has at least some basic ability to segment the larger components present in the samples, and showing the robustness of this CAE-based approach for segmenting real biomedical HSI data. In addition to gland segmentation, the algorithms appeared to differentiate other tissue sections. However, these are not shown in the resultant HE stains, making it challenging to discern whether i) there really are unique tissues present, and ii) whether their boundaries have been accurately segmented. Smaller scale morphological features (e.g. the lumen) may be elucidated by the use of a larger number of clusters as can be seen in Appendix \ref{sec:large_clust_num}. Though without a precise ground truth, it is not possible to evaluate how accurately these objects may have been segmented, or whether they are truly present in each location. Therefore, we focus our attention on large-scale morphological features instead. 

The segmentations appear distorted at the bottom and right edges. This is a consequence of how segmentation images are reconstructed (by overlapping tiles from left to right and a top to bottom fashion), and the size of each image patch. A patch placed in the last column will only be overlapped by the patch placed below it, while a patch placed in the final row will only be overlapped by the patch to its right. This produces block-like/smeared artefacts with thicknesses equal to the patch size at the boundaries that can be easily cropped away. Aside from these artefacts, the reconstruction procedure may also subtly translate the position of objects by an amount that appears to be related to the chosen step size. This was observed in the Indian Pines dataset experiment (Appendix \ref{sec:IP_res}), and may be easily corrected with additional cropping. This artefact also affected the segmentations produced from the synthetic HSIs mentioned in the previous section, and contributed to lowering the segmentation quality scores.

\section{Conclusion}
We have investigated the use of convolutional autoencoders to perform unsupervised segmentation of Raman/IR HSIs by grouping regions with similar spatial and spectral features. The architectures contain two components: an autoencoder module that mines features from HSI patches and encodes them in low dimensional latent vectors, and a clustering module that then groups similar latent vectors together (akin to grouping together image patches based on the similarity of their constituent features). All of this can be performed/learned in an end-to-end fashion, ensuring the feature mining and compression are optimised for the subsequent clustering step. With a simulation study, we have shown that this approach is particularly useful in cases where different regions of a tissue contain similar molecular contents but may be differentiated by the way their constituent spectra are arranged in space. We also showed how the end-to-end nature of the architecture can improve the quality of the segmentations compared to a strategy that performs the feature mining and clustering in completely separate steps.

To assess whether these approaches can produce segmentations in the face of a much larger feature set that would normally be encountered in real HSI data (as well as experimental artefacts), we tested three autoencoder architectures on real IR HSI data: a generic 2D convolutioanl autoencoder, a generic 3D convolutional autoencoder, and a 2D architecture inspired by the recently proposed UwU-net. All segmentations were comparable and had good correspondence with HE stains used as approximate ground truths, indicating that this approach can cope with the diverse set of features found in real biomedical HSI data and other confounding experimental factors. Despite the hypothesised advantages of the UwU-net inspired architecture, no significant qualitative differences were observed when compared to the segmentations produced with the generic 2D or 3D CAEs. However, a precise/quantitative comparison between each architecture was not possible given the approximate nature of the ground truth HE stains, which depict an adjacent tissue slice and therefore do not provide information about the exact morphology/boundaries of each relevant component within the HSI data. Therefore, the true extent of the benefits any particular architecture may provide is uncertain. Additionally, the number of samples used in this study is small, and does not represent the full range of different features/sample types that may be encountered in practice (e.g. no tumoral samples were used in this study). Therefore, there remains some degree of uncertainty as to the quality we could expect from using this approach on a wider selection of samples.

It may be possible to use alternative datasets with accurate ground truths to compare architecture performance, such as satellite images of geographical landscapes. However, there are two potential issues with this. The first is that it is not clear whether features residing in specific portions of the spectral bands are as relevant as they are in biomedical HSI data. Therefore, this dataset may not allow us to observe the expected benefit from using any particular architecture for segmenting biomedical HSI data. Secondly, the accuracy of the approach depends on the accuracy of the initialised k-means cluster centres. Poor accuracy has been reported on the Indian Pines dataset and others in \cite{nalepa2020unsupervised} and confirmed separately in our own experiment (see Appendix \ref{sec:IP_res}). Therefore, it is likely that these kinds of datasets are a poor choice for comparing the performance of our architectures.

Another broader limitation of using this approach is that it is most effective when segmenting regions whose defining features span the patch size. Therefore, it may not be suitable for segmenting tissue regions whose characteristic spatial distribution of molecules spans large distances. Secondly, as mentioned above, the accuracy of the initialised cluster centres strongly determines the quality of the output segmentation. Therefore, this approach may be less effective in cases where there is an imbalance in the representation of tissue sections (i.e. one tissue class occupies a significantly smaller area than the others) \cite{franti2018k}. Furthermore, the segmentations suffer from subtle smearing/translation artefacts. Though, these appear to be straightforward to correct by cropping the image using prior knowledge of the patch size and step size.

As it currently stands, the results obtained with each architecture are the first to demonstrate the robustness of the the CAE-driven saptio-spectral clustering approach to segment major tissue components from biomedical HSIs (or arguably any kind of experimentally acquired HSI dataset). Nevertheless, in future work we aim to more precisely quantify the advantages of each architecture by applying them to data with accurate ground truth segmentations, such as more realistic simulated phantoms, or other biomedical HSI datasets. One unexplored feature is the use of the uncertainty encoded in the soft cluster assignments to display information about segmentation quality. This could enable a more rigorous comparison of the effectiveness of different segmentation architectures, or the assessment of how the quality of segmentations produced by different samples may vary from region to region. 
\newpage
\appendices
\section{Using additional cluster groups}
\label{sec:large_clust_num}
\begin{figure}[!htp]
    \centering
    \includegraphics[width=.9\columnwidth]{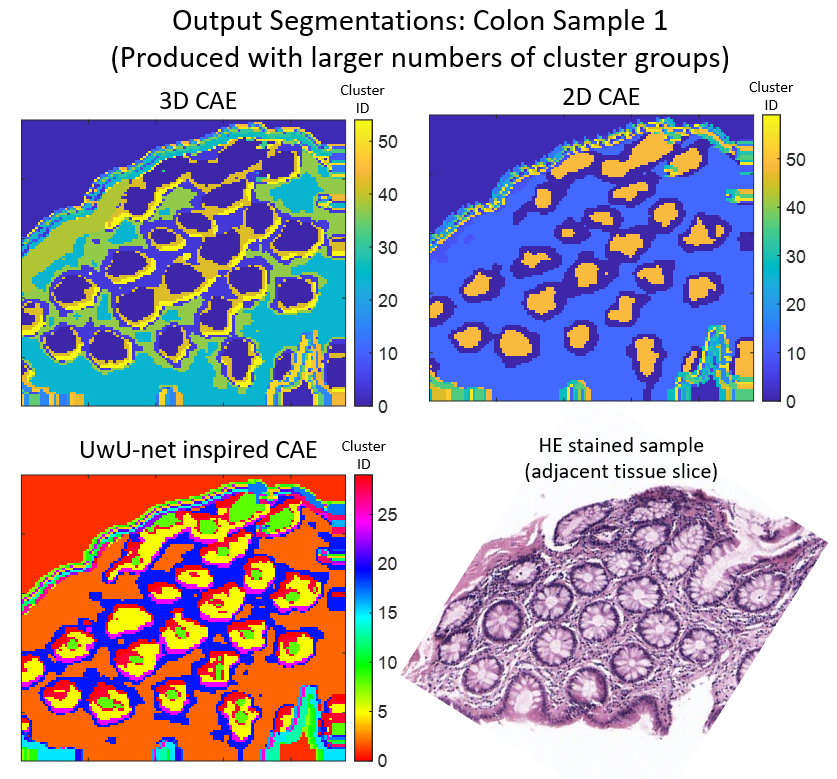}
    \caption{Segmentations for colon Sample 1, produced with a larger number of cluster groups.}
    \label{fig:more_clust}
\end{figure}

It may be possible to elucidate smaller scale morphological features from the colon images by using a larger number of clusters groups. E.g. it appears as though the lumen can be discerned from colon Sample 1 with the use of 40 cluster groups with the UwU-net inspired architecture (see Fig. \ref{fig:more_clust}). Though without a precise ground truth, it is not possible to assess how accurately these objects may have been segmented. The segmentations produced using additional cluster groups for the other two architecture types are shown as well.

\section{Other network architectures used in this work}
\label{sec:generic_cae}
\begin{figure*}[!htp]
    \centering
    \includegraphics[width=1.8\columnwidth]{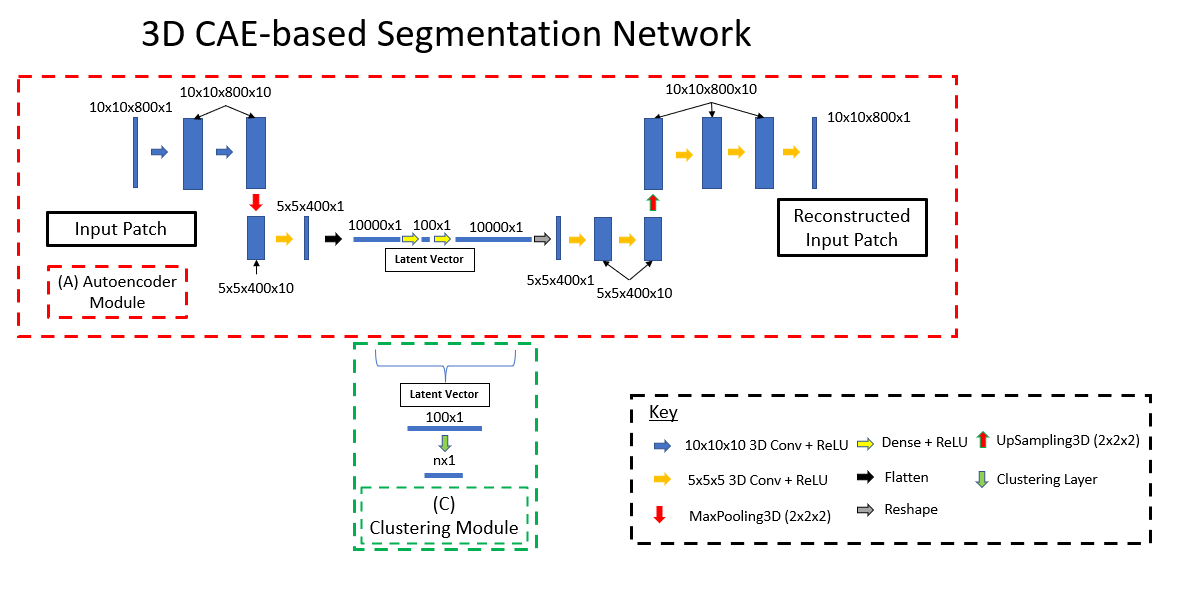}
    \caption{The generic 3D CNN-based architecture used to segment colon IR HSI images, where $C$ is a generic 3D CAE.}
    \label{fig:3dcnn}
\end{figure*}

\begin{figure*}[!htp]
    \centering
    \includegraphics[width=1.8\columnwidth]{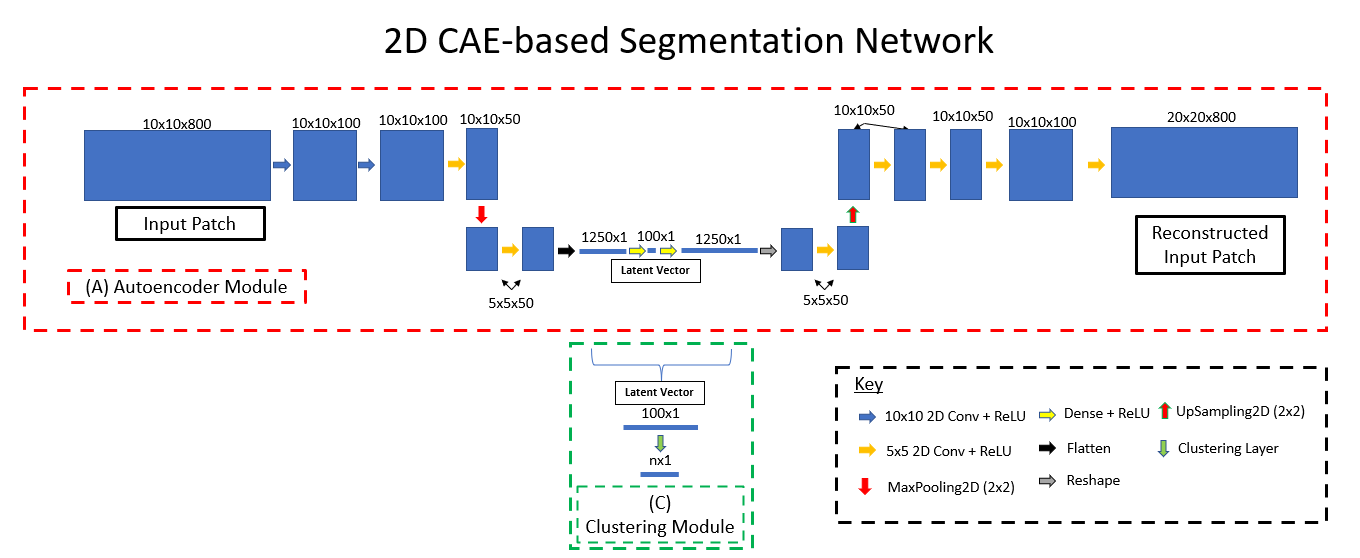}
    \caption{The generic 2D CNN-based architecture used to segment colon IR HSI images, where $C$ is a generic 2D CAE.}
    \label{fig:2dcnn}
\end{figure*}

\begin{figure*}[!htp]
    \centering
    \includegraphics[width=1.8\columnwidth]{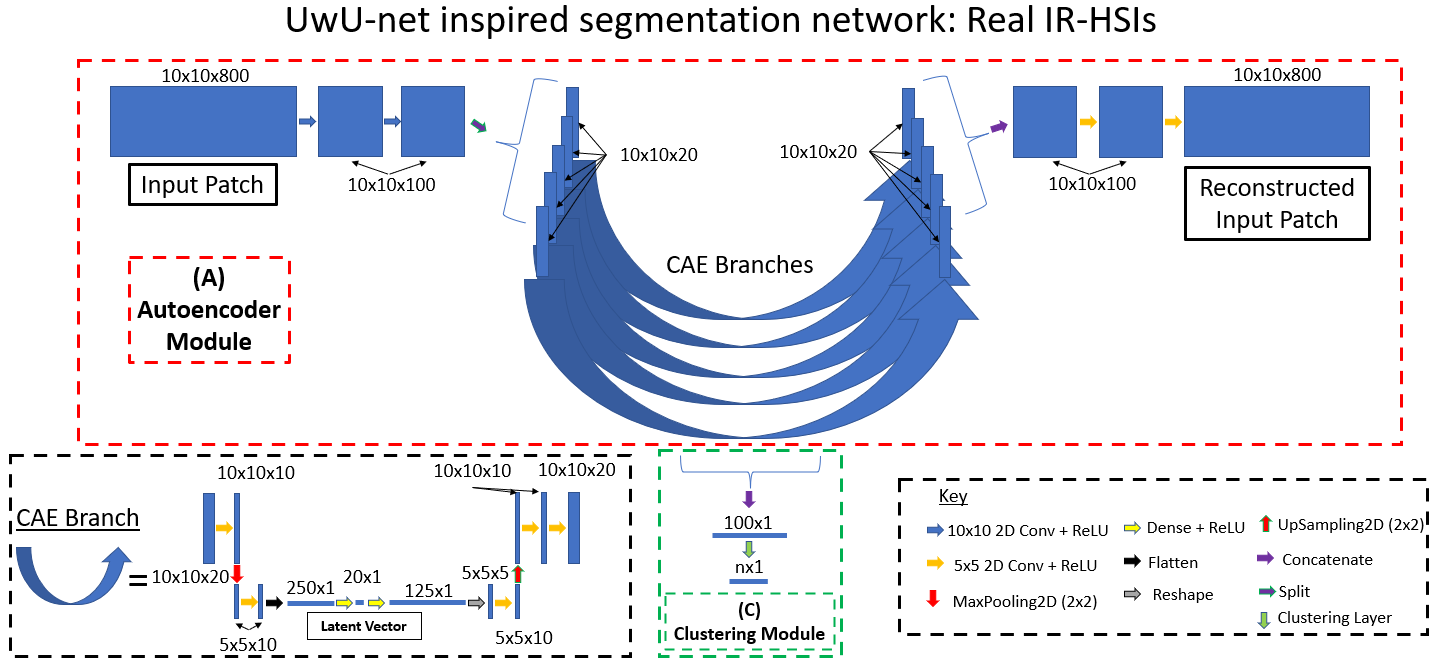}
    \caption{The UwU-net inspired architecture used to segment colon IR HSI images.}
    \label{fig:uwu_minerva}
\end{figure*}

In Section \ref{sec:real_HSI_method}, the segmentations acquired with the UwU-net inspired CAE architecture were compared to more generic CAE architectures based on 2D and 3D convolutional layers. Here we show all three architectures (Figs. \ref{fig:3dcnn}, \ref{fig:2dcnn}, and \ref{fig:uwu_minerva}). The 2D architecture featured a similar number of convolutional filters and layers as well as the same size latent vector as the UwU-net inspired architecture to ensure that any observed differences in segmentation quality would mostly be a consequence of the use of parallel CAEs. The 3D CAE featured fewer layers and filters to reduce computation time. As discussed in Section \ref{sec:real_HSI_method}, it is unclear how this may have affected the expressive power of this architecture relative to the 2D architectures given the 3D architecture is utilising information about 3D features. 
\newpage

\begin{landscape}
\section{CAE pretraining loss curves}
This section contains the loss curves produced at the pretraining stage of most of the networks used in this work.
\label{sec:loss_curves}
\begin{figure}[!htp]
    \centering
    \includegraphics[width=.8\columnwidth]{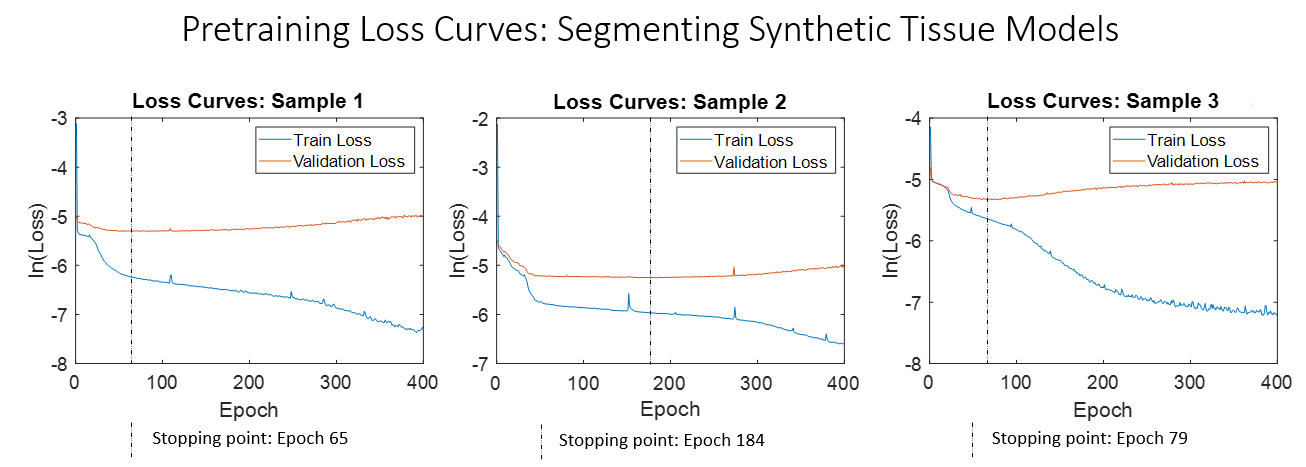}
    \caption{Validation and training losses for the pretraining of the UwU-net inspired CAE used to segment the synthetic Raman HSIs. The stopping points are shown with dashed lines.}
    \label{fig:synthetic_loss}
\end{figure}
\begin{figure}[!htp]
    \centering
    \includegraphics[width=.8\columnwidth]{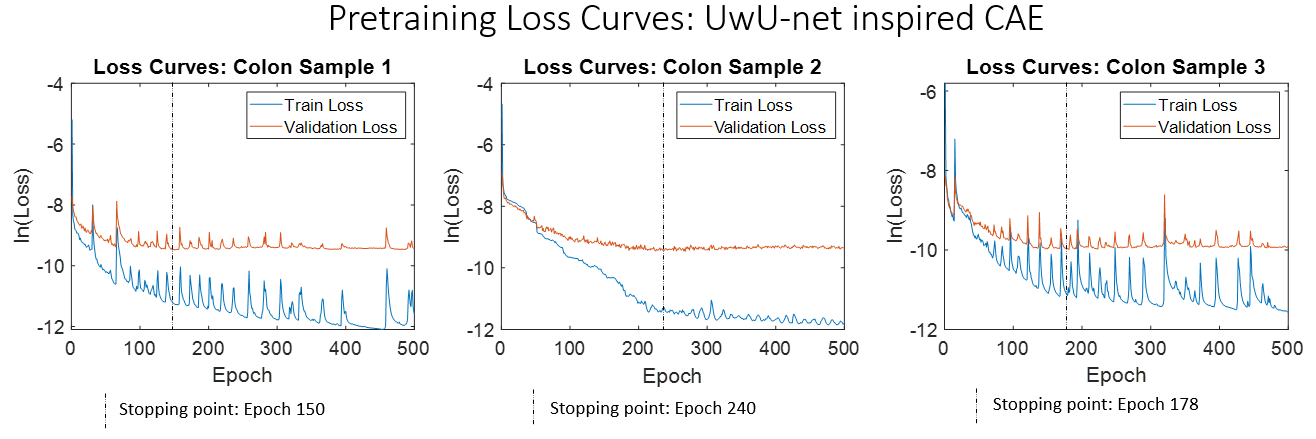}
    \caption{Validation and training losses for the pretraining of the UwU-net inspired CAE used to segment the colon IR-HSIs. The stopping points are shown with dashed lines.}
    \label{fig:minerva_loss_uwu}
\end{figure}
\end{landscape}
\begin{landscape}
\begin{figure}[!htp]
    \centering
    \includegraphics[width=.8\columnwidth]{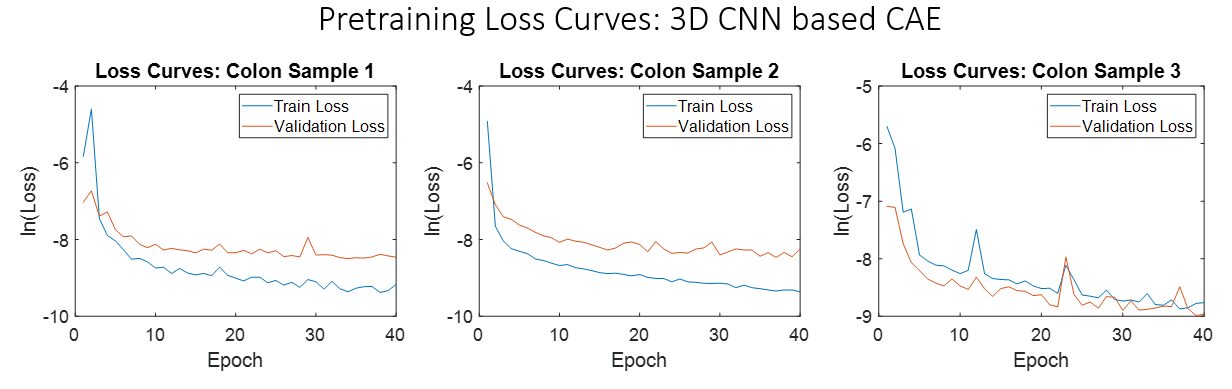}
    \caption{Validation and training losses for the pretraining of the 3D CAE used to segment the colon IR-HSIs.}
    \label{fig:minerva_loss_3d}
\end{figure}
\begin{figure}[!htp]
    \centering
    \includegraphics[width=.8\columnwidth]{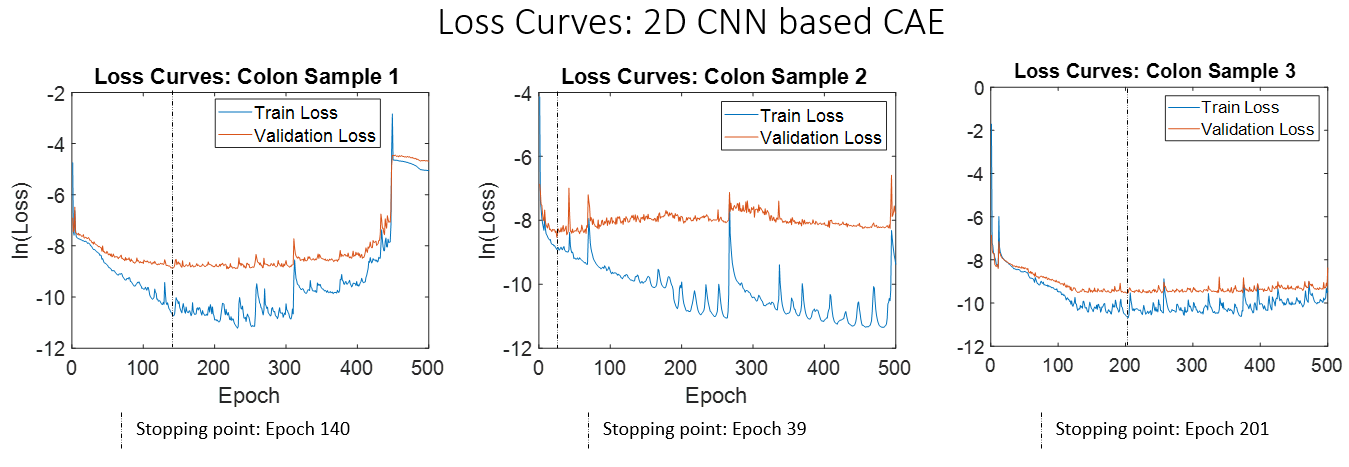}
    \caption{Validation and training losses for the pretraining of the generic 2D CAE used to segment the colon IR-HSIs. The stopping points are shown with dashed lines.}
    \label{fig:minerva_loss_2d}
\end{figure}
\end{landscape}
\section{Segmention of the Indian Pines dataset}
\label{sec:IP_res}
The Indian Pines dataset (IP) is often used to assess the performance of HSI segmentation algorithms \cite{nalepa2020unsupervised} (\url{https://www.ehu.eus/ccwintco/index.php/Hyperspectral_Remote_Sensing_Scenes}). It consists of a geographical landscape scene with 17 classes (including the background) acquired with 200 spectral reflectance bands (where 24 tied to water absorption have been removed) in the wavelength range $0.4 - 2.5 10^{-6}$ m. At first glance it would seem like a suitable dataset for assessing the performance of our segmentation algorithms. However, this dataset and similar landscape scene HSIs are possibly poor choices for a couple of reasons. The first is that when CAE-based spatio-spectral clustering approaches have been applied to these datasets, the segmentation produced at the initialisation stage has notably poor accuracy (as reported in \cite{nalepa2020unsupervised}). This is problematic, as the end-to-end clustering technique works on the principle that the most confident cluster assignments produced at the initialisation stage are accurate. This was confirmed using a UwU-net inspired CAE in our own work, presented below. Secondly, it is not clear whether the satellite HSIs contain the same kinds of features that would be relevant to segmenting biomedical HSIs (e.g. distinct, sharp peaks occurring in specific subsets of the spectral band). Therefore, it may not be suitable to assess the expected performance of each architecture when specifically applied to biomedical HSI data.

\begin{figure}[]
    \centering
    \includegraphics[width=.95\columnwidth]{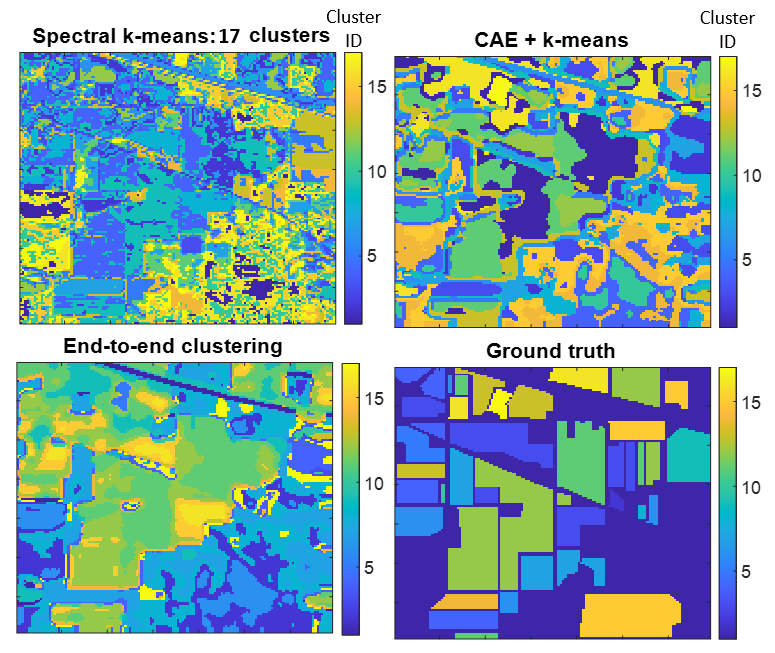}
    \caption{The results of applying our spatio-spectral clustering approach to the IP dataset (all images assigned randomly indexed colour scales). The resultant segmentation does not effectively differentiate all components of the image. This is verified quantitatively using an NMI score shown in Table \ref{tab:IP_res_tab}.}
    \label{fig:IP_res}
\end{figure}

\begin{table}[!htp]
\begin{center}
\label{tab:IP_res_tab}

\caption{Indian Pines Dataset Segmentation Accuracy}

\begin{tabular}{lllll}
  &  &  &   &  \\
    &  &  &   &  \\
NMI Score & CAE+k-means & End-to-end & Spectral k-means &  \\
          & 0.43              & 0.43             &  0.42 &
\end{tabular}

\begin{tabular}{lllll}
ARS Score & CAE+k-means & End-to-end & Spectral k-means &  \\
          & 0.23              & 0.21             &  0.21 &
\end{tabular}
\end{center}
\label{tab:IP_res_tab}
\end{table}

The following describes our own experiment using an UwU-net inspired convolutional autoencoder on the IP dataset. The dataset was cropped to have dimensions 140x140x200, and then upsampled to give it dimensions of 280x280x200 (to enable use of 10x10 patches for simplified downssampling without increasing spatial context encoded in each patch). The resulting image was decomposed into $10\times10$ patches in steps of 2 following the same procedure described in Section \ref{sec:data_prep_sim}. 12000 patches were used for the training set while the remaining 6496 were used for the validation set. Though, all patches were used in the final evaluation, meaning there was some degree of data leakage in this experiment. The network architecture was the same as that shown in Fig. \ref{fig:uwu_minerva}. The autoencoder module was initially trained for 500 epochs with a batch size of 80 (though, not to optimality, as the validation loss was never observed to consistently rise - however, it also showed very little change during the later stages of training, so it is expected that the model is fairly optimised). Then, both the autoencoder and clustering module were trained for 10 updates, or until the number of changed cluster assignments was less than $0.1\%$. The resultant segmentation was downsized back to 120x120 and cropped to account for the tiling offsetting the segmentation in the $x$ and $y$ dimensions. It is apparent that the resultant segmentation does not accurately differentiate most of the regions within the HSI. The accuracy of the segmentation was evaluated quantitatively using an NMI score, and ARS. Background pixels (whose locations were determined using the ground truth) were not used in the calculation of these metrics. In both cases, spatio-spectral clustering outperformed k-means but both scores are low. The NMI scores were comparable to those reported in \cite{nalepa2020unsupervised}.

\section*{Acknowledgments}

The authors acknowledge support from the UKRI EPSRC grant EP/V047914/1 : Terabotics – terahertz robotics for surgery and medicine.

\bibliography{bibliography.bib}

\ifCLASSOPTIONcaptionsoff
  \newpage
\fi

\end{document}